\documentclass{PoS}

\title{Chiral behaviour of matrix elements of $\Delta B = 2$ and $\Delta C = 2$
operators\thanks{Work supported by Taiwanese National Science Council via 
grants NSC96-2112-M-009-020-MY3, NSC96-2119-M-007-001, and U.S. Department of Energy 
via grant DE-FG03/974014.}}

\ShortTitle{Chiral behaviour of $\Delta B = 2$ and $\Delta C =2$ matrix elements}

\author{William~Detmold\\
        Department of Physics, University of Washington, Seattle, WA 98155, U.S.A.\\
        E-mail: \email{wdetmold@phys.washington.edu}}

\author{\speaker{C.-J.~David~Lin}\\
        Institute of Physics, National Chiao-Tung University, Hsinchu 300, Taiwan\\
        Physics Division, National Centre for Theoretical Sciences, Hsinchu 300, Taiwan\\
        E-mail: \email{dlin@mail.nctu.edu.tw}}

\abstract{We investigate the light-quark mass and spatial volume dependence of matrix elements
of $\Delta B = 2$ and $\Delta C = 2$ four-fermion operators.  These operators are relevant for 
$B^{0}_{(s)}{-}\bar{B}^{0}_{(s)}$ and $D^{0}{-}\bar{D}^{0}$ mixing in and beyond the Standard
Model.  An important conclusion of this work 
is that the chiral
extrapolations for matrix elements of heavy-light meson mixing beyond
the Standard Model are more complicated than that for the Standard Model
mixing matrix elements.}

\FullConference{The XXV International Symposium on Lattice Field Theory\\
		 July 30-4 August 2007\\
		 Regensburg, Germany}

\def\op{{\mathcal{O}}}

\begin{document}
\section{Introduction}
A precise calculation for neutral $B$ mixing matrix elements has been an urgent task
since the recent measurement of $\Delta m_{s}$ \cite{Abulencia:2006ze}.  Such a calculation
is crucial in obtaining stringent constraints on the unitarity triangle of
the Cabibbo-Kobayashi-Maskawa (CKM) matrix and searching for new physics.  On the other hand, 
the $D^{0}{-}\bar{D}^0$ mixing system is a good channel to search for new physics
\cite{Bergmann:2000id}, because the Standard Model contribution is
strongly suppressed.  In the Standard Model, the short distance contribution to
the mass differences of these heavy neutral meson mixing systems
is predominantly determined by the matrix
elements of a single set of four quark operators:
\begin{eqnarray}
  \label{eq:1}
   {\cal O}_{1,aa} &=& \bar{h}^{\alpha} \gamma_{\mu} (1-\gamma_{5})
   q_{a}^{\alpha} 
           \mbox{ }\bar{h}^{\beta} \gamma_{\mu} (1-\gamma_{5}) q_{a}^{\beta} ,
\end{eqnarray}
where $h$ is a heavy quark field (either a $b$ or a $c$ quark),
$q_{a}$ is a light-quark field with flavour $a$ ($a$ is not summed
over), and $\alpha$ and $\beta$ are colour indices.  Models containing
flavour-changing currents other than the $V-A$ form (arising in
supersymmetric extensions of the Standard Model and other scenarios)
usually result in mass differences that additionally depend on matrix
elements of the four-quark operators~\cite{Gabbiani:1996hi}
\begin{eqnarray}
 {\cal O}_{2,aa} &=& \bar{h}^{\alpha} (1-\gamma_{5}) q_{a}^{\alpha}
           \mbox{ }\bar{h}^{\beta} (1-\gamma_{5}) q_{a}^{\beta} ,
\nonumber\\
\label{eq:quark_level_op_susy} 
 {\cal O}_{3,aa} &=& \bar{h}^{\alpha} (1-\gamma_{5}) q_{a}^{\beta}
           \mbox{ }\bar{h}^{ \beta} (1-\gamma_{5}) q_{a}^{\alpha} ,
\\
 {\cal O}_{4,aa} &=& \bar{h}^{\alpha}  (1-\gamma_{5}) q_{a}^{\alpha}
           \mbox{ }\bar{h}^{\beta}   (1+\gamma_{5}) q_{a}^{\beta} ,
\nonumber\\
 {\cal O}_{5,aa} &=& \bar{h}^{\alpha}  (1-\gamma_{5}) q_{a}^{\beta}
           \mbox{ }\bar{h}^{\beta}  (1+\gamma_{5}) q_{a}^{\alpha} ,
\nonumber
\end{eqnarray}
Generically we can
represent these operators as
\begin{eqnarray}
  \label{eq:2}
  \op_{i,aa}= \bar{h} \Gamma_1 q \, \bar{h} \Gamma_2 q \,,
\end{eqnarray}
for the appropriate choice of spin and colour matrices,
$\Gamma_{1,2}$.  In lattice calculations, it is convenient to perform
a Fierz transformation which renders linear combinations of the
operators in Eq.~(\ref{eq:quark_level_op_susy}) into products of
colour-singlet currents. Nevertheless, we choose to work in the basis of
Eq.~(\ref{eq:quark_level_op_susy}).

In this article, we present a study for the light-quark mass 
and spatial volume dependence in matrix elements relevant to the mass differences in
the neutral $B$ and $D$ meson mixing systems, using (partially quenched) heavy
meson chiral perturbation theory\footnote{Some of the matrix elements studied in this
work are also relevant to the width difference in the $B$ mixing 
system~\cite{Lenz:2006hd}.}.  So far, only one exploratory numerical calculation in 
quenched lattice QCD has been carried out for the full set of these matrix 
elements~\cite{Becirevic:2001xt}.  However, with experimental progress in CKM
physics, an accurate determination of these matrix elements, which involves reliable chiral
extrapolations, will become necessary in the foreseeable
future.  In this work, we show that the chiral extrapolation
for matrix elements of ${\cal O}_{1,aa}$ is considerably less
complicated than that for matrix elements of the operators in
Eq.~(\ref{eq:quark_level_op_susy}).  This is due to the fact 
${\cal O}_{1,aa}$ preserves heavy quark spin symmetry, as explained in
Sections \ref{sec:4q} and \ref{sec:matrix_elements}.
Details of the formulae for the chiral extrapolations of these
matrix elements have been presented in Ref.~\cite{Detmold:2006gh}.

\section{Heavy meson chiral perturbation in finite volume}
\label{sec:hmchpt}
Heavy meson chiral perturbation theory (HM$\chi$PT) was formulated
in Refs.~\cite{Burdman:1992gh,Wise:1992hn,Yan:1992gz}, and generalised
to the quenched and partially-quenched versions in 
Refs.~\cite{Booth:1994hx,Sharpe:1995qp}.  Finite spatial volume effects
in this effective theory have also been investigated in Ref.~\cite{Arndt:2004bg}.
Here we only sketch the 
ingredients necessary to study neutral heavy-light meson mixing systems.
Details of this effective theory can be found in the above references.

In HM$\chi$PT,
the heavy-light meson field appears in the covariant form
\begin{equation}
\label{eq:HQ_field}
 H^{(Q)}_{a} = \frac{1 + \slash\!\!\! v}{2} \left ( 
 P^{\ast (Q)}_{a,\mu} \gamma^{\mu} - P^{(Q)}_{a}\gamma_{5}\right ) ,
\end{equation}
where $P^{(Q)}_{a}$ and $P^{\ast (Q)}_{a,\mu}$ annihilate
pseudoscalar and vector mesons containing a heavy quark $Q$ and a
light anti-quark of flavour $a$. In the heavy particle formalism, such
mesons have momentum $p^\mu=M_P v^\mu +k^\mu$ with $|k^\mu|\ll M_P$
and $v^\mu$ is the velocity of the particle.  Under a heavy quark spin
$SU(2)$ transformation $S$ and a generic light-flavour transformation
$U$ [{\it i.e.}, $U\in SU(3)$ for full QCD and $U\in SU(6|3)$ for
partially-quenched QCD (PQQCD)],
\begin{equation}
\label{eq:HQ_field_transformation}
 H^{(Q)}_{a} \longrightarrow S H^{(Q)}_{b} U^{\dagger}_{ba} .
\end{equation}
The conjugate field, which creates heavy-light mesons containing a
heavy quark $Q$ and a light anti-quark of flavour $a$, is defined as
\begin{equation}
\label{eq:barHQ_field}
 \bar{H}^{(Q)}_{a} = \gamma^{0} H^{(Q)\dagger} \gamma_{0}
 = \left ( 
 P^{\ast (Q)\dagger}_{a,\mu} \gamma^{\mu} + 
    P^{(Q)\dagger}_{a}\gamma_{5}\right )
 \frac{1 + \slash\!\!\! v}{2} ,
\end{equation}
which transforms under $S$ and $U$ as
\begin{equation}
\label{eq:barHQ_field_transformation}
 \bar{H}^{(Q)}_{a} \longrightarrow U_{ab} \bar{H}^{(Q)}_{b} S^{\dagger} .
\end{equation}
These heavy-light meson fields couple to the Goldstone meson fields
\begin{equation}
\label{eq:xi_field}
 \xi \equiv {\mathrm e}^{i\Phi/f} = \sqrt{\Sigma} ,
\end{equation}
where $\Sigma$ is the ordinary nonlinear Goldstone field.
The field $\xi$ transforms as
\begin{equation}
\label{eq:xi_field_transformation}
 \xi \longrightarrow U_{\mathrm{L}} 
  \xi U^{\dagger} = U \xi 
   U^{\dagger}_{\mathrm{R}},
\end{equation}
where $U_{\mathrm{L(R)}}$ is an element of the left-handed
(right-handed) $SU(3)$ and $SU(6|3)$ groups for QCD and PQQCD
respectively.  

The mesons containing a heavy anti-quark
$\bar{Q}$ and a light quark of flavour $a$ can be included in the 
theory by applying the
charge conjugation operation to the above heavy-light meson field $H^{(Q)}_{a}$.
The field that annihilates such mesons is
\begin{equation}
 \label{eq:HQbar_field}
 H^{(\bar{Q})}_{a} = \left ( 
 P^{\ast (\bar{Q})}_{a,\mu} \gamma^{\mu} - P^{(\bar{Q})}_{a}\gamma_{5}\right ) 
 \frac{1 - \slash\!\!\! v}{2},
\end{equation}
which transforms under $S$ and $U$ as
\begin{equation}
\label{eq:HQbar_field_transformation}
 H^{(\bar{Q})}_{a} \longrightarrow U_{ab} H^{(\bar{Q})}_{b} S^{\dagger} .
\end{equation}

The main effects of heavy quark symmetry breaking in HM$\chi$PT Lagrangian
is the splitting between vector and pseudoscalar heavy-light meson
masses~\cite{Boyd:1994pa} 
\begin{equation}
\label{eq:HQ_spin_breaking_term}
\frac{\lambda_{2}}{M_{P}}
{\mathrm{tr_{D}}} \left (
 \bar{H}^{(Q)}_{a}\sigma_{\mu\nu} H^{(Q)}_{a} \sigma^{\mu\nu}
\right ) ,
\end{equation}
where $\lambda_{2}$ is a low energy constant (LEC), and ${\mathrm{tr_{D}}}$
means trace in the spinor indices.

\section{Four-fermion operators}
\label{sec:4q}
Under a chiral transformation, the four-quark operators in
Eq.~(\ref{eq:quark_level_op_susy}) fall into two categories:
\begin{eqnarray}
 \op_{LL} &=& \bar{h}\mbox{ }\Gamma_{LL}\mbox{ }q_{L} 
 \mbox{ }\mbox{ }\bar{h}\mbox{ }\Gamma_{LL}\mbox{ }q_{L} ,
\nonumber\\
\label{eq:two_categories}
 \op_{LR} &=& \bar{h}\mbox{ }\Gamma^{(1)}_{LR}\mbox{ }q_{L} 
 \mbox{ }\mbox{ }\bar{h}\mbox{ }\Gamma^{(2)}_{LR}\mbox{ }q_{R} ,
\end{eqnarray}
where
\begin{equation}
 q_{L,R} = \frac{1\pm\gamma_{5}}{2} q .
\end{equation}

Operators $\op_{1,aa},\,\op_{2,aa}$ and $\op_{3,aa}$ are of the
first type and transform in the symmetric $({\bf 6_L,1_R})$
representation built from the direct product $({\bf
  3_L,1_R})\otimes({\bf 3_L,1_R})=({\bf 6_L,1_R})\oplus({\bf
  \overline{3}_L,1_R})$ under chiral rotations while $\op_{4,aa}$ and
$\op_{5,aa}$ are of the second type and transform in the $({\bf
  3_L,3_R})$ representation. Here we refer to the SU(3) flavour
transformation properties, leaving the partially quenched extension to
the following subsection. Note that the colour indices in
Eq.~(\ref{eq:quark_level_op_susy}) are relevant to short-distance
physics, and hence play no role in the chiral properties of these
operators.  Treating $\Gamma_{LL}$,
$\Gamma^{(1)}_{LR}$ and $\Gamma^{(2)}_{LR}$ as spurions transforming
as
\begin{eqnarray}
 \Gamma_{LL} \longrightarrow S\mbox{ }\Gamma_{LL}\mbox{ }U^{\dagger}_{L} ,
\nonumber\\
 \Gamma^{(1)}_{LR} \longrightarrow S\mbox{ }
    \Gamma^{(1)}_{LR}\mbox{ }U^{\dagger}_{L} ,
\nonumber\\
 \Gamma^{(2)}_{LR} \longrightarrow S\mbox{ }
   \Gamma^{(2)}_{LR}\mbox{ }U^{\dagger}_{R} ,
\end{eqnarray}
the operators in Eq.~(\ref{eq:two_categories}) remain invariant under
heavy-quark spin and chiral rotations.  
The bosonisation of the operators in Eqs.~(\ref{eq:1}) and (\ref{eq:quark_level_op_susy})
can be performed using these spurion transformation properties.
The procedure is explained in details in Ref.~\cite{Detmold:2006gh}. It leads to the following
set of operators involving the individual heavy meson fields:
\begin{eqnarray}
 \op^{\rm HM\chi PT}_{1,aa} &=&  \beta_{1} \left [ 
            \left ( \xi P^{(h)\dagger} \right )_{a}
            \left ( \xi P^{(\bar{h})} \right )_{a}
          + \left ( \xi P^{\ast (h)\dagger}_{\mu} \right )_{a}
            \left ( \xi P^{\ast (\bar{h}),\mu} \right )_{a}
                 \right ] \,,
\nonumber\\
\label{eq:tree_level_ops}
 \op^{\rm HM\chi PT}_{2(3),aa} &=&  \beta_{2(3)} 
            \left ( \xi P^{(h)\dagger} \right )_{a}
            \left ( \xi P^{(\bar{h})} \right )_{a} 
            +  \beta_{2(3)}^\prime
            \left ( \xi P^{\ast (h)\dagger}_{\mu} \right )_{a}
            \left ( \xi P^{\ast (\bar{h}),\mu} \right )_{a}\,,
\\
 \op^{\rm HM\chi PT}_{4(5),aa} &=&  \beta_{4(5)}  
            \left ( \xi P^{(h)\dagger} \right )_{a}
            \left ( \xi^{\dagger} P^{(\bar{h})} \right )_{a} 
           + \hat\beta_{4(5)}
            \left ( \xi^{\dagger} P^{(h)\dagger} \right )_{a}
            \left ( \xi P^{(\bar{h})} \right )_{a} 
\nonumber\\
&&           + \beta_{4(5)}^\prime
            \left ( \xi P^{\ast (h)\dagger}_{\mu} \right )_{a}
            \left ( \xi^{\dagger} P^{\ast (\bar{h}),\mu} \right )_{a}
           + \hat\beta_{4(5)}^\prime
            \left ( \xi^{\dagger} P^{\ast (h)\dagger}_{\mu} \right )_{a}
            \left ( \xi P^{\ast (\bar{h}),\mu} \right )_{a}\,,
\nonumber
\end{eqnarray}
where the $\beta_i,\, \beta_i^\prime,\hat\beta_i$, and
$\hat\beta_i^\prime$ are LECs.

It is important to note that in the above equation, the operator
${\cal O}^{\rm HM\chi PT}_{1,aa}$ behaves somewhat differently from
the other operators as only a single LEC, $\beta_{1}$, occurs.  This
greatly simplifies any chiral extrapolation of corresponding lattice
data for neutral heavy-light meson mixing in the Standard Model, as
discussed in the next section. We
stress that this simplification is not obvious from the operator
structure and is particular to the $V-A$ form
of the Standard Model currents.  In general, one would expect
that operators for
pseudoscalar and vector meson mixing processes are accompanied by
different LECs.  This is the case for all the
non-Standard-Model operators, as shown in
Eq.~(\ref{eq:tree_level_ops}).

To understand the origin of the above simplification in the Standard
Model operator $\op^{\rm HM\chi PT}_{1,aa}$, we turn to heavy quark
effective theory (HQET).
In this effective
theory, the operators that produce the same matrix elements as those
in Eq.(\ref{eq:2}) are \cite{Flynn:1990qz}
\begin{eqnarray}
  \label{eq:5}
  \op_{i,aa}^{\rm HQET} &=& \tilde{Q}\Gamma_1 q_a\, Q^\dagger \Gamma_2 q_a
  + Q^\dagger \Gamma_1 q_a \, \tilde{Q}\Gamma_2 q_a\,, 
\end{eqnarray}
where $\Gamma_{1,2}$ are the appropriate Dirac and colour structures
from Eq.~(\ref{eq:quark_level_op_susy}). Here, $Q$ and $\tilde{Q}$
denote fields annihilating a heavy quark and heavy anti-quark,
respectively (these fields do not create the corresponding
anti-particles).  Additional terms in HQET which create two heavy
quarks or annihilate two heavy anti-quarks will not contribute to
neutral heavy-meson mixing and are ignored.

The standard model operator in HQET, $\op_{1,aa}^{\rm HQET}$,
satisfies the relation
\begin{eqnarray}
  \label{eq:6}
  \left\{S_Q^3, \op_{1,aa}^{\rm HQET} \right\} | P \rangle 
= \left\{S_Q^3, \op_{1,aa}^{\rm HQET} \right\} | \bar{P} \rangle
&=& 0\,,
\end{eqnarray}
where $|P\rangle$ is pseudoscalar heavy-light meson state, and
$|\bar{P}\rangle$ is the state of its anti-particle.  The operator
\begin{equation}
\label{eq:HQ_spin_op}
 S_Q^3=\epsilon^{ij3}[Q^\dagger \sigma_{ij}Q - \tilde{Q}
\sigma_{ij} \tilde{Q}^\dagger] ,
\end{equation}
is the heavy quark spin operator \cite{Savage:1990di} that changes the
spin of the heavy-light meson state by one. Therefore,
Eq.~(\ref{eq:6}) implies that the mixing matrix elements for vector
and pseudoscalar heavy-light mesons are equal and opposite
in the heavy-quark limit.  This symmetry is
reflected in HM$\chi$PT, leading to the result for $\op^{\rm HM\chi
  PT}_{1,aa}$ in Eq.~(\ref{eq:tree_level_ops}).

For the non-Standard-Model operators, it is straightforward to show
that
\begin{equation}
\label{eq:non_sm_ops_anti_comm}
  \left\{S_Q^3, \op_{i,aa}^{\rm HQET} \right\} | P \rangle \not= 0 , \, \,
  \left\{S_Q^3, \op_{i,aa}^{\rm HQET} \right\} | \bar{P} \rangle \not= 0 ,
\end{equation}
and
\begin{equation}
\label{eq:non_sm_ops_comm}
  \left [ S_Q^3, \op_{i,aa}^{\rm HQET} \right ] | P \rangle \not= 0 , \, \,
  \left [ S_Q^3, \op_{i,aa}^{\rm HQET} \right ] | \bar{P} \rangle \not= 0 ,
\end{equation}
where $i=2,3,4,5$. This means that the pseudoscalar and vector meson
mixing processes via these operators are not proportional to each
other, hence the appearance of the terms accompanied by
$\beta^{\prime}_{2,3,4,5}$ and $\hat{\beta}^{\prime}_{4,5}$ in
Eq.~(\ref{eq:tree_level_ops}).

We end this section by noting that
equations of motion for the heavy quark~\cite{Becirevic:2001xt} result in
$\op_{3,aa} =-\op_{1,aa}/2-\op_{2,aa}$, and can be used to relate
some of the LECs in Eq.~(\ref{eq:tree_level_ops}).

\section{Neutral meson mixing matrix elements}
\label{sec:matrix_elements}

\begin{figure}[!t]
  \centering
\includegraphics[width=0.35\columnwidth]{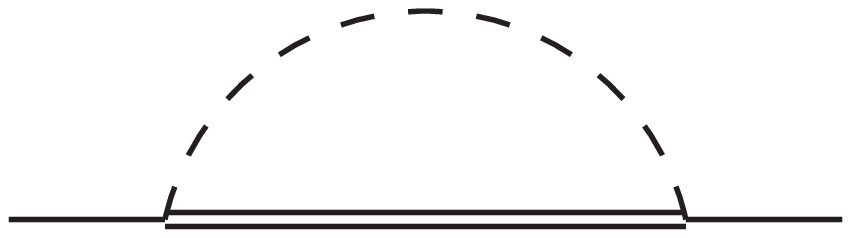}
\hspace*{1mm}
\includegraphics[width=0.225\columnwidth]{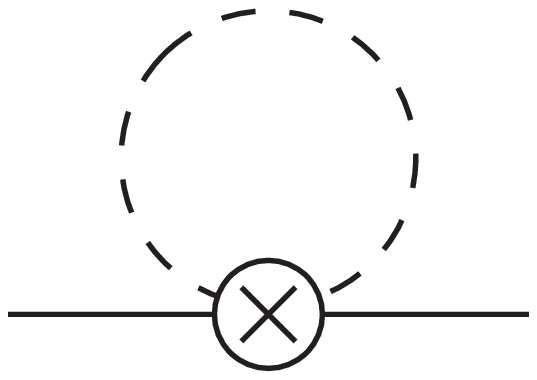}
\hspace*{1mm}
\includegraphics[width=0.35\columnwidth]{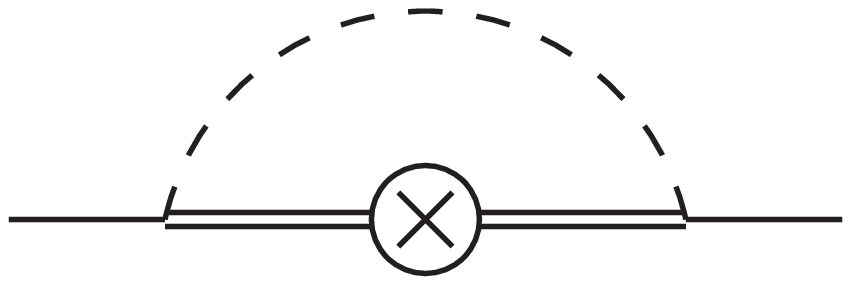} 
\\
\hspace*{0.2cm}(a)\hspace*{4.2cm}(b)\hspace*{4.2cm}(c)
  \caption{Diagrams contributing to the matrix elements of four-quark
    operators at NLO in the chiral expansion. Solid, double and dashed
    lines correspond to propagators of pseudoscalar and vector
    heavy-light mesons, and Goldstone mesons, respectively. The
    crossed circle denotes the four-quark operator and diagram (a) is
    the wave-function renormalisation.}
  \label{fig:oneloop}
\end{figure}

Calculations at NLO in the chiral expansion require the evaluation of
the one-loop diagrams shown in Figure~\ref{fig:oneloop}. We perform
these calculations both at infinite volume and in a cubic spatial box
of dimensions $L^3$ (the time extent is assumed to be infinite). In
this section we summarise the results, relegating details of the
calculations to Ref.~\cite{Detmold:2006gh}.  Although we
present results specifically in the $B$-meson systems, note that they
are also applicable to $D$-meson systems, under the assumption that
the charm-quark mass is large enough compared to $\Lambda_{\rm QCD}$.

For the standard model operator we find the following matrix elements
\begin{eqnarray}
  \langle \bar{B}^0 | \op_{1,dd}|B^0\rangle &=&
  \beta_1\left(1+ {\cal T}^{(1)}_{d} + \frac{{\cal 
        W}_{\bar{B}^0}+{\cal W}_{B^0}}{2}+{\cal
      Q}^{(1)}_{d} \right) + {\rm analytic \, \, terms}\,, 
\nonumber
\\
  \label{eq:7}
\\
  \langle \bar{B}^0_s | \op_{1,ss}|B^0_s\rangle &=&
  \beta_1\left(1+ {\cal T}^{(1)}_{s} + \frac{{\cal
        W}_{\bar{B}^0_s}+{\cal W}_{B^0_s}}{2}+{\cal
      Q}^{(1)}_{s} \right) + {\rm analytic \, \, terms}\, .
\nonumber
\end{eqnarray}
The ``analytic terms'' here include Goldstone meson mass squared
terms, a term $\sim \alpha_s(M_b)/4\pi$ (arising from mixing) and a
term $\sim \Lambda_{\rm QCD}/M_b$. The wave-function contributions,
${\cal W}_M$ (diagram (a) in Figure~\ref{fig:oneloop}), 
and tadpole- and sunset-type operator renormalisations,
${\cal T}^{(i)}_{a}$ and ${\cal Q}^{(i)}_{a}$ (diagrams (b) and (c) in
Figure~\ref{fig:oneloop}, respectively) are non-analytic functions of
the light quark mass and lattice volume and are defined in
Ref.~\cite{Detmold:2006gh}.

For the operators that contribute to the $B$-meson mixing
processes beyond the Standard Model, we obtain:
\begin{eqnarray}
  \langle \bar{B}^0 | \op_{2(3),dd}|B^0\rangle &=&
  \beta_{2(3)}\left(1+ {\cal T}^{(2(3))}_{d} + \frac{{\cal 
        W}_{\bar{B}^0}+{\cal W}_{B^0}}{2} \right) + 
\beta_{2(3)}^\prime {\cal Q}^{(2(3))}_{d} + \ldots\,, 
\nonumber
\\
\nonumber
\\
  \langle \bar{B}^0_s | \op_{2(3),ss}|B^0_s\rangle &=&
  \beta_{2(3)}\left(1+ {\cal T}^{(2(3))}_{s} + \frac{{\cal 
        W}_{\bar{B}^0_s}+{\cal W}_{B^0_s}}{2} \right) + 
\beta_{2(3)}^\prime {\cal Q}^{(2(3))}_{s} + \ldots\,. 
\nonumber 
\\
 \label{eq:10}
\\
  \langle \bar{B}^0 | \op_{4(5),dd}|B^0\rangle &=&
  \left[\beta_{4(5)}+\hat\beta_{4(5)}\right] 
  \left(1+ {\cal T}^{(4(5))}_{d} + \frac{{\cal 
        W}_{\bar{B}^0}+{\cal W}_{B^0}}{2} \right) + 
\left[\beta_{4(5)}^\prime+\hat\beta_{4(5)}^\prime\right] 
 {\cal Q}^{(4(5))}_{d} +\ldots\,, 
\nonumber
\\
\nonumber 
\\
  \langle \bar{B}^0_s | \op_{4(5),ss}|B^0_s\rangle &=&
  \left[\beta_{4(5)}+\hat\beta_{4(5)}\right] 
\left(1+ {\cal T}^{(4(5))}_{s} + \frac{{\cal 
        W}_{\bar{B}^0_s}+{\cal W}_{B^0_s}}{2} \right) + 
\left[\beta_{4(5)}^\prime+\hat\beta_{4(5)}^\prime\right]
 {\cal Q}^{(4(5))}_{s} +\ldots\, ,
\nonumber
\end{eqnarray}
where ``$\ldots$'' represents the ``analytic terms'' as in Eq.~(\ref{eq:7}).
The terms $\sim{\cal Q}_q^{(i)}$ arising from the sunset diagrams
[Fig.~\ref{fig:oneloop}(c)] involve the neutral heavy-light vector
meson mixing amplitudes.  As discussed in the preceding section, it is
only in the case of ${\cal O}_{1,qq}$ that these amplitudes are
related to those of the pseudoscalar heavy-light mesons. For
$i=2,3,4,5$ these terms are consequently accompanied by different LECs.

The above results lead to the fact that the chiral extrapolations
for $B$ mixing matrix elements in and beyond the Standard Model
have different features.  Generically, the chiral expansion
for $\langle \bar{B}^{0}_{(s)} | {\cal O}_{1,aa} | B^{0}_{(s)}\rangle$
takes the form
\begin{equation}
\label{eq:O_1_chiral_generic}
 \langle \bar{B}^{0}_{(s)} | {\cal O}_{1,aa} | B^{0}_{(s)}\rangle
 \stackrel{{\rm chiral}}{\longrightarrow}
 \gamma_{1} \left ( 1 + L \right ) + {\rm analytic}\mbox{ }\mbox{ }{\rm terms},
\end{equation}
where $\gamma_{1}$ is the leading-order LEC, $L$
denotes the non-analytic one-loop contributions (chiral logarithms),
and the analytic terms are from the next-to-leading-order
counter-terms in the chiral expansion.  However, for the operators in
Eq.~(\ref{eq:quark_level_op_susy}), the chiral expansion has the
generic feature:
\begin{equation}
\label{eq:O_2345_chiral_generic}
 \langle \bar{B}^{0}_{(s)} | {\cal O}_{i,aa} | B^{0}_{(s)}\rangle
 \stackrel{{\rm chiral}}{\longrightarrow}
 \gamma_{i} \left ( 1 + L \right ) + \gamma^{\prime}_{i} L^{\prime}
  + {\rm analytic}\mbox{ }\mbox{ }{\rm terms},
\end{equation}
where $i=2,3,4,5$, $\gamma_{i}$ and $\gamma^{\prime}_{i}$ are unknown
leading-order LECs, and $L$ and $L^{\prime}$ are
different one-loop chiral logarithms. 
The appearance of the second non-analytic term complicates the chiral 
extrapolation in Eq.~(\ref{eq:O_2345_chiral_generic}) 
because an additional unknown parameter must be determined.

We end this article by noting that the effects of scalar resonances 
in these matrix elements have been studied in Ref.~\cite{Becirevic:2006me} and found
to be negligible.


\begin{thebibliography}{99}
%
\bibitem{Abulencia:2006ze}
  A.~Abulencia {\it et al.}  [CDF Collaboration],
  Phys.\ Rev.\ Lett.\  {\bf 97} (2006) 242003
  [arXiv:hep-ex/0609040].
%
\bibitem{Bergmann:2000id}
  S.~Bergmann, Y.~Grossman, Z.~Ligeti, Y.~Nir and A.~A.~Petrov,
  Phys.\ Lett.\  B {\bf 486} (2000) 418
  [arXiv:hep-ph/0005181].
%
\bibitem{Gabbiani:1996hi}
  F.~Gabbiani, E.~Gabrielli, A.~Masiero and L.~Silvestrini,
  Nucl.\ Phys.\  B {\bf 477} (1996) 321
  [arXiv:hep-ph/9604387].

%
\bibitem{Lenz:2006hd}
  A.~Lenz and U.~Nierste,
  JHEP {\bf 0706} (2007) 072
  [arXiv:hep-ph/0612167].

%
\bibitem{Becirevic:2001xt}
  D.~Becirevic, V.~Gimenez, G.~Martinelli, M.~Papinutto and J.~Reyes,
  JHEP {\bf 0204} (2002) 025
  [arXiv:hep-lat/0110091].
%
\bibitem{Detmold:2006gh}
  W.~Detmold and C.-J.~D.~Lin,
  Phys.\ Rev.\  D {\bf 76} (2007) 014501
  [arXiv:hep-lat/0612028].
%
\bibitem{Burdman:1992gh}
  G.~Burdman and J.~F.~Donoghue,
  Phys.\ Lett.\  B {\bf 280} (1992) 287.
%
\bibitem{Wise:1992hn}
  M.~B.~Wise,
  Phys.\ Rev.\  D {\bf 45} (1992) 2188.
%
\bibitem{Yan:1992gz}
  T.~M.~Yan, H.~Y.~Cheng, C.~Y.~Cheung, G.~L.~Lin, Y.~C.~Lin and H.~L.~Yu,
  Phys.\ Rev.\  D {\bf 46}, 1148 (1992)
  [Erratum-ibid.\  D {\bf 55}, 5851 (1997)].
%
\bibitem{Booth:1994hx}
  M.~J.~Booth,
  Phys.\ Rev.\  D {\bf 51}, 2338 (1995)
  [arXiv:hep-ph/9411433].
%
\bibitem{Sharpe:1995qp}
  S.~R.~Sharpe and Y.~Zhang,
  Phys.\ Rev.\  D {\bf 53}, 5125 (1996)
  [arXiv:hep-lat/9510037].
%
\bibitem{Arndt:2004bg}
  D.~Arndt and C.~J.~D.~Lin,
  Phys.\ Rev.\  D {\bf 70} (2004) 014503
  [arXiv:hep-lat/0403012].
%
\bibitem{Boyd:1994pa}
  C.~G.~Boyd and B.~Grinstein,
  Nucl.\ Phys.\  B {\bf 442} (1995) 205
  [arXiv:hep-ph/9402340].
%
\bibitem{Flynn:1990qz}
  J.~M.~Flynn, O.~F.~Hernandez and B.~R.~Hill,
  Phys.\ Rev.\  D {\bf 43}, 3709 (1991).
%
\bibitem{Savage:1990di}
  M.~J.~Savage and M.~B.~Wise,
  Phys.\ Lett.\  B {\bf 248}, 177 (1990).
%
\bibitem{Becirevic:2006me}
  D.~Becirevic, S.~Fajfer and J.~Kamenik,
  JHEP {\bf 0706}, 003 (2007)
  [arXiv:hep-ph/0612224].
%
\end{thebibliography}
\end{document}